\documentclass{ws-procs975x65}            
\begin{document}
\title{
Higgs Mass in D-Term triggered Dynamical  SUSY Breaking
}

\author{Nobuhito Maru}

\address{Department of Mathematics and Physics, Osaka City University \\
Osaka, 558-8585, Japan \\
}

\begin{abstract}
We discuss a new mechanism of D-term dynamical supersymmetry breaking in the context of Dirac gaugino scenario. 
The existence of a nontrivial solution of the gap equation for D-term is shown. 
It is also shown that an observed 126 GeV Higgs mass is realized by tree level D-term effects in a broad range of parameters. 
\end{abstract}


\bodymatter

\section{Introduction}

Supersymmetry (SUSY) is one of the leading candidates of 
physics beyond the Standard Model (SM) solving the hierarchy problem. 
Unfortunately, we have no indication of SUSY at present. 
Furthermore, an observed Higgs boson mass of 126 GeV discovered at LHC puts severe constraints on 
 the minimal supersymmetric standard model (MSSM) parameter space, namely the MSSM with light sparticles. 
In this situation, we have two approaches to proceed. 
One is the MSSM with heavy sparticles, the so-called high scale SUSY breaking scenario 
 which gives up to solve the naturalness problem. 
The other is extensions of the MSSM. 
We take the latter approach and focus on the Dirac gaugino scenario \cite{DG} as one of the many interesting extensions.  
The setup to realize Dirac gaugino scenario is as follows. 
The gauge sector in the MSSM is extended to ${\cal N}=2$ matter content  
 by introducing chiral superfields in the adjoint representations of the SM gauge group, 
$\Phi_a=(\phi_a, \psi_a, F_a)~(a=SU(3), SU(2), U(1))$. 
The matter sector is ${\cal N}=1$. 
Dirac gaugino masses are generated from the dimension five operator
\begin{eqnarray}
{\cal L} = \int d^2 \theta \sqrt{2} \frac{{\cal W}^0_\alpha {\cal W}^\alpha_a \Phi_a}{\Lambda}
= \frac{\langle D^0 \rangle}{\Lambda} \psi_a \lambda_a + \cdots, 
\label{supersoft}
\end{eqnarray}
if a $U(1)$ D-term vacuum expectation value (VEV) 
 $\langle D^0 \rangle \subset {\cal W}_\alpha^0 = \theta_\alpha D^0$ in the hidden sector is nonzero. 
${\cal W}^0_\alpha, {\cal W}^a_\alpha$ represents the field strength tensor superfield 
 for the hidden $U(1)$ and the SM gauge groups, respectively. 
$\Lambda$ is a cutoff scale. 

Sfermion masses are generated at 1-loop, 
\begin{eqnarray}
M^2_{\tilde{f}} \simeq \frac{C_a(f)\alpha_a}{\pi} M_{\lambda_a}^2 \log \left( \frac{m_{\phi_a}^2}{M^2_{\lambda_a}} \right) 
\end{eqnarray}
where $C_a(f)$ is a group theoretical factor. 
$\alpha_a$ denotes fine structure constant for the SM gauge coupling constant. 
$M_{\lambda_a}, m_{\phi_a}$ are gaugino and adjoint scalar masses.  
These masses are flavor-blind since it is induced by the SM gauge interactions. 
Therefore, we have no SUSY flavor and CP problems. 

In Dirac gaugino scenario, gaugino masses are typically heavier than the sfermion masses by a factor $4\pi/g$. 
This fact relaxes the bounds from LHC experiment for spectrum since the gluino and squark production are suppressed 
 comparing to the Majorana gaugino case. 
 
\section{A New Mechanism of D-term Dynamical SUSY Breaking}

In this section, we discuss a new mechanism of D-term dynamical SUSY breaking 
 generating mixed Majorana-Dirac gaugino masses, 
 which we proposed several years ago \cite{IMaru1}. 

Consider a SUSY $U(N)$ gauge theory with adjoint chiral superfield $\Phi^a$. 
The Lagrangian is given by 
\begin{eqnarray}
{\cal L} = \int d^4\theta K(\Phi^a, \overline{\Phi}^a, V) 
+ \int d^2 \theta {\rm Im}\frac{1}{2} {\cal F}_{ab}(\Phi^a) {\cal W}^{a \alpha}{\cal W}_\alpha^a 
+ \left[
\int d^2\theta W(\Phi^a) + {\rm h.c.}
\right]  
\label{Lag}
\end{eqnarray} 
where $K, {\cal F}$ and $W$ are K\"ahler potential, the gauge kinetic function which is holomorphic in $\Phi^a$, 
 and $W$ is a superpotential whose $F$-term is assumed to be vanished. 
The subscripts $a, b$ of ${\cal F}$ denote the derivative with respect to $\Phi^{a,b}$. 
We regard an overall $U(1)$ in $U(N)$ gauge group as the hidden gauge group 
 and denotes the components of the hidden $U(1)$ fields with superscript or subscript ``0".   

Focusing on the fermion mass terms derived from eq.(\ref{Lag}), we find
\begin{eqnarray}
-\frac{1}{2} 
(\lambda^a, \psi^a)
\left(
\begin{array}{cc}
0 & -\frac{\sqrt{2}}{4} {\cal F}_{ab0} D^0 \\
-\frac{\sqrt{2}}{4} {\cal F}_{ab0} D^0 & \partial_a \partial_b W \\
\end{array}
\right)
\left(
\begin{array}{c}
\lambda^b \\
\psi^b \\
\end{array}
\right). 
\end{eqnarray}
The mass eigenvalues of (seesaw-type) mass matrix are easily obtained 
\begin{eqnarray}
m_{\pm} = \frac{1}{2} \langle \partial_a \partial_a W \rangle 
\left[
1 \pm \sqrt{1 + \left( \frac{2\langle D \rangle}{\langle \partial_a \partial_a W \rangle} \right)^2} 
\right], 
\quad D \equiv -\frac{\sqrt{2}}{4} {\cal F}_{0aa} D^0
\end{eqnarray}
if $\langle D \rangle \ne 0$ and $\langle \partial_a \partial_a W \rangle \ne 0$. 
Gaugino masses ($m_-$) becomes massive by nonzero $D^0$ and SUSY is spontaneously broken. 

D-term equation of motion tells us that the D-term VEV is given 
 by Dirac bilinear condensate between the gaugino and the adjoint fermion, 
\begin{eqnarray}
\langle D^0 \rangle \sim \langle {\cal F}_{0ab} \psi^a \lambda^b  
+ \bar{{\cal F}}_{0ab} \bar{\psi}^a \bar{\lambda}^b \rangle.  
\end{eqnarray}
As in the BCS and the NJL models, 
the D-term VEV is determined by solving the gap equation for D-term. 

Our strategy of the potential analysis is as follows. 
Note that we have three constant background fields $\phi^0, D^0, F^0$. 
We simply work in a region where $\langle D^0 \rangle \gg \langle F^0 \rangle$, 
 namely F-term is perturbative comparing to D-term. 
Some sample points satisfying $\langle D^0 \rangle \gg \langle F^0 \rangle$ were numerically found in \cite{IMaru2}. 
We first solve the stationary conditions 
\begin{eqnarray}
&&\frac{\partial V(D, \phi, \bar{\phi}, F=\bar{F}=0)}{\partial D}=0, 
\label{stD} \\
&&\frac{\partial V(D, \phi, \bar{\phi}, F=\bar{F}=0)}{\partial \phi}=\frac{\partial V(D, \phi, \bar{\phi}, F=\bar{F}=0)}{\partial \bar{\phi}}=0. 
\label{stphi}
\end{eqnarray}
Then, we find the stationary values $(D_*, \phi_*, \bar{\phi}_*)$. 
Using these stationary values, the stationary conditions for F-term is solved perturbatively,   
\begin{eqnarray}
\frac{\partial V(D_*, \phi_*, \bar{\phi}_*, F, \bar{F})}{\partial F}
=\frac{\partial V(D_*, \phi_*, \bar{\phi}_*, F, \bar{F})}{\partial \bar{F}}=0
\label{stF}
\end{eqnarray}
and the stationary values of F-term $(F_*, \bar{F}_*)$ are found. 
Note that the stationary condition for D-term (\ref{stD}) is nothing but the gap equation which we wish to solve. 

The 1-loop effective potential is calculated 
\begin{eqnarray} 
V = N^2 |m_\phi|^4 
\left[
c_1(\phi, \bar{\phi}) \Delta_0^2 
+ \frac{1}{32\pi^2} 
\left(
c_2 \Delta_0^4 -|\lambda^{(+)}|^4 \log |\lambda^{(+)}|^2 
- |\lambda^{(-)}|^4 \log |\lambda^{(-)}|^2 \right)
\right] \nonumber \\
\end{eqnarray}
where $\lambda^{(\pm)} \equiv m_{\pm}/\langle \partial_a \partial_a W \rangle = \frac{1}{2}(1 \pm \sqrt{1+\Delta_0^2}), 
\Delta_0 \equiv 2\langle D \rangle/\langle \partial_a \partial_a W \rangle$. 
$c_1$ is some function of $\phi$ and $\bar{\phi}$. 
$c_2$ is a constant \cite{IMaru3}. 

We can immediately obtain the gap equation for D-term from the stationary condition of D-term.  
\begin{eqnarray}
0 = \frac{\partial V}{\partial D}
&=& \Delta_0 
\left[
c_1 + \frac{1}{64\pi^2} 
\left\{
4c_2 \Delta_0^2 
\right. \right. \nonumber \\
&& \left. \left. 
-\frac{1}{\sqrt{1+\Delta_0^2}} 
\left(
\lambda^{(+)3} (2 \log \lambda^{(+)2} +1 )
- \lambda^{(-)3} (2 \log \lambda^{(-)2} +1 )
\right)
\right\}
\right]. 
\label{gapeq}
\end{eqnarray}
Note that the gap equation has a trivial solution $\Delta_0=0$, which represents a SUSY vacuum. 
In Fig.\ref{fig1} (a), the plot of the quantity $\partial V/(\Delta_0 \partial D)$ as a function of $\Delta_0$ is shown. 
From this plot, we can find a nontrivial solution representing SUSY breaking vacuum.  
\def\figsubcap#1{\par\noindent\centering\footnotesize(#1)}
\begin{figure}[h]%
\begin{center}
  \parbox{2.3in}{\includegraphics[width=2in]{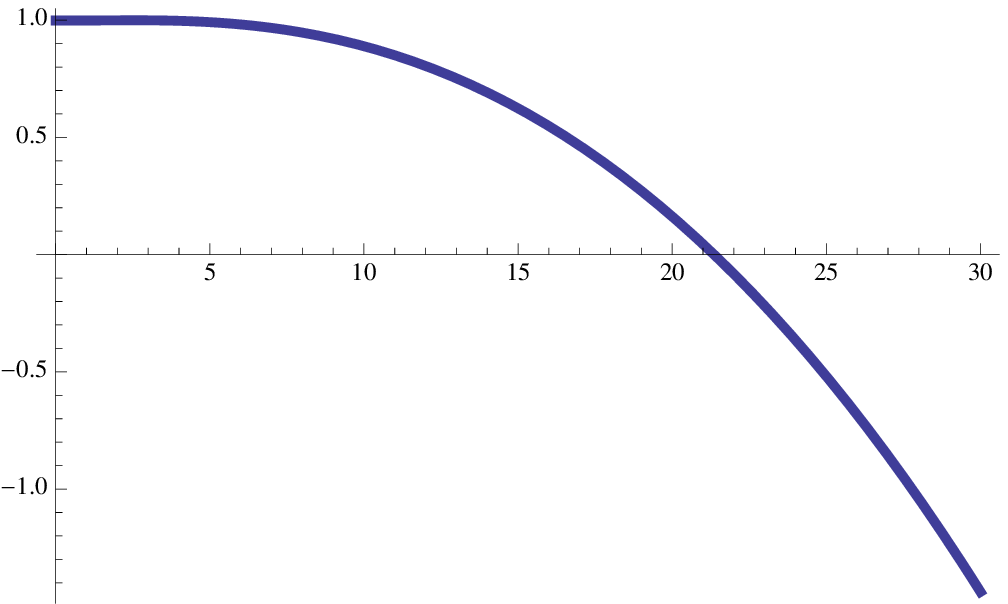}\figsubcap{a}}
  \hspace*{4pt}
  \parbox{1.8in}{\includegraphics[width=2in]{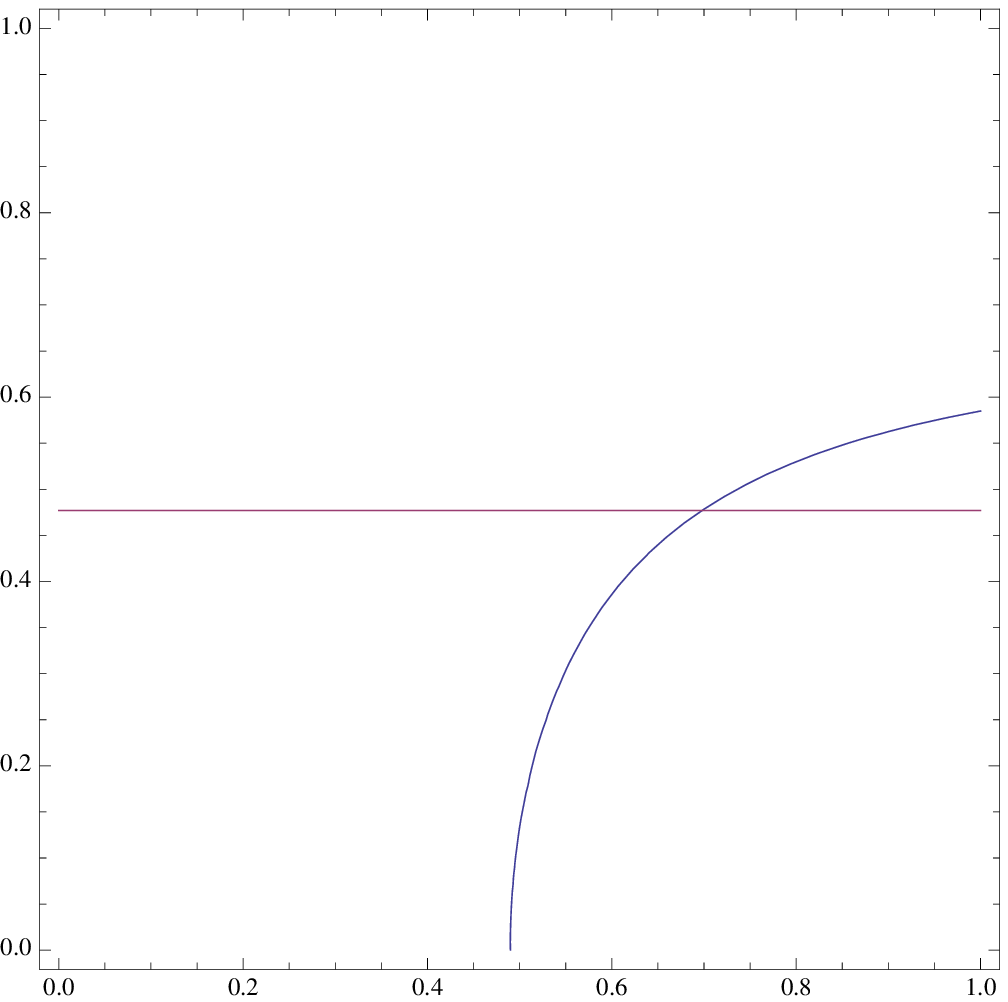}\figsubcap{b}}
  \caption{(a) The plot of the quantity $\partial V/(\Delta_0 \partial D)$ as a function of $\Delta_0$. 
  (b) The schematic picture of the intersection of the two curves 
  representing the gap equation (the straight line) and the $\phi$ flat condition (the curved one). 
  The horizontal (vertical) axis is $\phi/\Lambda (\Delta_0)$. 
  }%
  \label{fig1}
\end{center}
\end{figure}
The stationary value of the adjoint scalar fields 
 is determined by the intersection point of the two curves representing the gap equation and the stationary condition for $\phi$ 
 in the $(\Delta_0, \phi=\bar{\phi})$ plane.  
The schematic picture is shown in Fig.\ref{fig1} (b). 

The SUSY breaking vacuum cannot be a global minimum of the potential 
 since the energy of SUSY theory is positive semi-definite and 
 a SUSY minimum with the vanishing energy is a solution of the gap equation as mentioned above.  
We have to check whether our SUSY breaking is sufficiently long-lived 
 by taking into account a tunneling effect from our false SUSY breaking vacuum to the true SUSY one. 
Its tunneling rate can be roughly estimated and should be small
\begin{eqnarray}
\exp 
\left[
-\frac{\langle \Delta \phi \rangle^4}{\langle \Delta V \rangle} 
\right]
\sim 
\exp 
\left[
-\frac{(\Delta_0 \Lambda)^2}{m_\phi^2} 
\right] 
\ll1
\end{eqnarray} 
where $\Delta \phi$ is a field distance between the two vacua, which is roughly given by $\Delta_0\Lambda$. 
$\Delta V$ is a potential height between them, which is given by D-term squared $(m_\phi \Delta_0 \Lambda)^2$. 
In order for our false vacuum to be sufficiently long-lived, $m_\phi \ll \Delta_0 \Lambda$ is required, 
 which can be always satisfied 
 since the adjoint scalar mass is a free superpotential mass parameter.

\section{Higgs Mass via D-term Effects}

In this section, we discuss whether an observed 126 GeV Higgs mass can be realized in our framework \cite{IMaru3}. 
The Lagrangian of Higgs sector is 
\begin{eqnarray}
{\cal L}_{{\rm Higgs}} &=& \int d^4 \theta
 \left[
 H_u^\dag e^{-g_Y V_1 - g_2 V_2 - 2q_u g V_0} H_u
 +H_d^\dag e^{g_Y V_1 - g_2 V_2 - 2q_d g V_0} H_d
 \right]  \nonumber \\
&& + \left[ \left(  \int d^2 \theta \mu H_u \cdot H_d \right)
- B\mu H_u \cdot H_d + {\rm h.c.} \right].
\label{HiggsLagrangian}
\end{eqnarray}
We have adopted notation $X \cdot Y \equiv \epsilon_{AB} X^A Y^B = X^A Y_A = - Y \cdot X$,
$\epsilon_{12} = - \epsilon_{21} = \epsilon^{21} = - \epsilon^{12} =1$.
$V_{1,2,0}$ are vector superfields of the SM gauge group and that of the overall $U(1)$ respectively
 and the corresponding gauge couplings are denoted by $g_{Y,2}$ and $g$, respectively.
Unlike the MSSM case, the soft scalar Higgs masses $m_{H_u}^2 |H_u|^2, m_{H_d}^2 |H_d|^2$ are not introduced
 since they are induced by $D$-term contributions in our framework.

The Higgs potential is derived from (\ref{HiggsLagrangian}) 
\begin{eqnarray}
V_{{\rm Higgs}} &=& \frac{g_2^2}{2(1+{\rm Im}{\cal F}'''\langle \Phi_0 \rangle)}
\left( H_u^\dag \frac{\sigma^a}{2} H_u + H_d^\dag \frac{\sigma^a}{2} H_d \right)^2 
\nonumber \\
&&
+ \frac{g_Y^2}{8(1+{\rm Im}{\cal F}'''\langle \Phi_0 \rangle)} \left( |H_u|^2 - |H_d|^2 \right)^2
\nonumber \\
&&
+ \frac{1}{2(1+{\rm Im}{\cal F}'''\langle \Phi_0 \rangle)}
\left( q_u g |H_u|^2 + q_d g |H_d|^2 - \left. \frac{\partial \Gamma^{{\rm 1-loop}}(D^0)}{\partial D^0} \right|_{D^0=D^{0*}} \right)^2
\nonumber \\
&&
+ |\mu|^2 (|H_u|^2 + |H_d|^2) + (B \mu H_u \cdot H_d + {\rm h.c.}).
\end{eqnarray}
Here we have denoted by $D^{0*}$ the solution to the improved gap equation, 
\begin{eqnarray}
0=(1+{\rm Im}{\cal F}'''\langle \Phi_0 \rangle) D^0 -q_u g |H_u|^2 -q_d g |H_d|^2 
+ \frac{\partial \Gamma^{{\rm 1-loop}}(D^0)}{\partial D^0}. 
\end{eqnarray}
The deviation $\delta D^{0*}$ of the value from $D^{0*}$ in \cite{IMaru3} is in fact small
by the ratio of electroweak scale and SUSY breaking scale.
Therefore, we approximate the solution to the improved gap equation
 by the value of $D^{0*}$ in \cite{IMaru3} denoted as $\langle D^0 \rangle$.
Taking into account the fact that ${\rm Im}{\cal F}'''\langle \Phi_0 \rangle \sim \langle \Phi_0 \rangle/\Lambda \ll 1$,
 we neglect the term ${\rm Im}{\cal F}'''\langle S \rangle$ at the leading order.
The resulting Higgs potential at the leading order is given by
\begin{eqnarray}
V_{{\rm Higgs}}
&\simeq&
\frac{g_2^2 + g_Y^2}{8}
\left[
|H_u^0|^2 - |H_d^0|^2
\right]^2
+ \frac{1}{2}
\left(
q_u g |H_u^0|^2 + q_d g |H_d^0|^2 - \langle D^0 \rangle
\right)^2 \nonumber \\
&&+ |\mu|^2 \left( |H_u^0|^2 + |H_d^0|^2\right)
- \left( B \mu H_u^0 H_d^0 + {\rm h.c.} \right)
\nonumber \\
&=&
\frac{g_2^2 + g_Y^2}{32} v^4 c_{2\beta}^2
+ \frac{v^2}{2}
\left[
\mu^2 - B \mu s_{2\beta}
\right]
+ \frac{1}{8}
\left(
( q_u g s_\beta^2 + q_d g c_\beta^2)v^2 - 2 \langle D^0 \rangle
\right)^2
\end{eqnarray}
where we have restricted the potential to the CP-even neutral sector
of Higgs doublets $H_u=(H_u^+, H_u^0)^T$, $H_d=(H_d^0, H_d^-)^T$. 
In the last line,
 the neutral components of Higgs fields are parametrized as
\begin{eqnarray}
&&H_u^0 = \frac{1}{\sqrt{2}} \left[ s_\beta (v+h) + c_\beta H + i (c_\beta A - s_\beta G^0 ) \right], \\
&&H_d^0 = \frac{1}{\sqrt{2}} \left[ c_\beta (v+h) - s_\beta H + i (s_\beta A + c_\beta G^0 ) \right]
\end{eqnarray}
where
$s_\beta \equiv \sin \beta, c_\beta \equiv \cos \beta$. 
$h, H, A$ are the SM Higgs, the CP even heavy Higgs and CP odd Higgs, respectively. 
$G^0$ is the would-be Nambu-Goldstone boson eaten as the longitudinal component of $Z$-boson.
The VEV of Higgs field is $v \simeq 246$~GeV and $\frac{g_Y^2+g_2^2}{4}v^2=M_Z^2$ in this convention.

We are now ready to calculate Higgs mass.
As in the MSSM,
 the minimization of the scalar potential $\partial V_{{\rm Higgs}}/\partial v^2 = \partial V_{{\rm Higgs}}/\partial \beta = 0$
 allows us to express $\mu$ and $B\mu$ in terms of other parameters.
\begin{eqnarray}
\mu^2 + \frac{M_Z^2}{2} &=& 
\frac{g}{2c_{2\beta}} \left( (q_u s_\beta^2 + q_d c_\beta^2 )g v^2 - 2 \langle D^0 \rangle \right)
\left(
q_u s_\beta^2 - q_d c_\beta^2
\right), \\
M_A^2 \equiv \frac{2B \mu}{s_{2\beta}}
&=& 2 \mu^2 
+ \frac{q_u+q_d}{2}g \left( (q_u s_\beta^2 + q_d c_\beta^2 ) gv^2 - 2 \langle D^0 \rangle \right) \nonumber \\
&=& -M_Z^2 + \frac{q_u-q_d}{2 c_{2\beta}}g
\left( (q_u s_\beta^2 + q_d c_\beta^2 ) gv^2 - 2 \langle D^0 \rangle \right).
\end{eqnarray}
The Higgs mass is obtained as
\begin{eqnarray}
m_{{\rm Higgs}}^2 &=&
\frac{1}{2} \left[
\tilde{M}_Z^2 + M_A^2
-\sqrt{\left(
\tilde{M}_Z^2 + M_A^2
 \right)^2
 -4 \tilde{M}_Z^2 M_A^2c_{2\beta}^2
 }
\right]
\label{Hmass}
\end{eqnarray}
where
$\tilde{M}^2_Z \equiv M_Z^2  + q_u^2 g^2 v^2$.
It is interesting to see the correspondence 
 between our expression of Higgs mass (\ref{Hmass}) and that in the MSSM,
\begin{eqnarray}
m_{{\rm MSSM~Higgs}}^2 &=&
\frac{1}{2} \left[
M_Z^2 + M_A^2
-\sqrt{\left( M_Z^2 + M_A^2 \right)^2
 -4 M_Z^2 M_A^2c_{2\beta}^2 }
\right].
\end{eqnarray}
As in the case of MSSM, the upper bound of Higgs mass can be obtained
 by taking a decoupling limit $M_A^2 \to \infty$,
\begin{eqnarray}
m^2_{{\rm Higgs}} \to \tilde{M}_Z^2 c_{2\beta}^2.
\end{eqnarray}
$\tilde{M}_Z$ can be large enough by taking ${\cal O}(1)$ charge and coupling $q_u g$
\begin{eqnarray}
\tilde{M}_Z \sim \sqrt{(90~{\rm GeV})^2 + \left(246~{\rm GeV} \right)^2} \sim 262~{\rm GeV}.
\end{eqnarray}
Note that the minimization conditions of Higgs potential with $q_u+q_d=0$ to allow $\mu$-term in the superpotential leads to
\begin{eqnarray}
M_Z^2 + M_A^2 = -\frac{q_u g}{c_{2\beta}} \left( c_{2\beta} q_u g v^2 + 2 \langle D^0 \rangle \right).
\label{minimization}
\end{eqnarray}
In order to satisfy this condition,
 the dominant part in the right-hand side of (\ref{minimization})
 $q_u g \langle D^0 \rangle/c_{2\beta}$ is required to be negative.

Using this condition, we can eliminate $M_A^2$ in Higgs mass (\ref{Hmass}).
\begin{eqnarray}
m^2_{{\rm Higgs}} &=& \frac{1}{2}
\left[
-\frac{2q_u g}{c_{2\beta}} \langle D^0 \rangle
- \sqrt{\left(-\frac{2q_u g}{c_{2\beta}} \langle D^0 \rangle \right)^2
+ 8 c_{2\beta} q_u g \tilde{M}_Z^2 \langle D^0 \rangle + 4c_{2\beta}^2 \tilde{M}^4_Z}
\right]. 
\end{eqnarray}

A plot for 126 GeV Higgs mass as a function of $\tan \beta$ and $q_u g$ is shown below.
Here we have taken $q_u g > 0$ and $\cos 2\beta<0$ to satisfy the condition 
$q_u g \langle D^0 \rangle/c_{2\beta}<0$.
We can immediately see that 126 GeV Higgs mass is realized in a broad range of parameters.
Also, we found that the result is insensitive to the values of $D$-term VEV.
This fact is naturally expected from the non-decoupling nature of Higgs mass.

\begin{figure}[h]
\begin{center}
\includegraphics[width=2.0in]{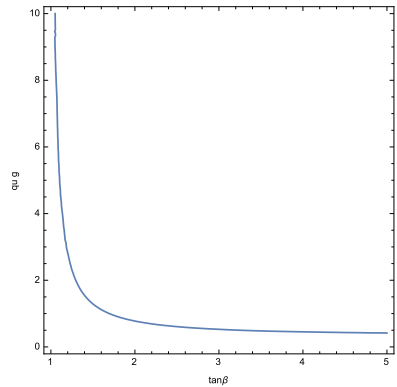}
\end{center}
\caption{A plot for the 126 GeV Higgs mass as a function of $\tan \beta$ and $q_u g$.}
\label{aba:fig1}
\end{figure}

\section{Summary}

Dirac gaugino scenario is an interesting extension of the MSSM, 
 which relaxes LHC constraints for the SUSY breaking parameters 
 by gluino and squark production suppression 
 comparing to Majorana gaugino case. 
Dirac gaugino mass is generated by the nonvanishing D-term VEV. 
We have proposed a new mechanism of D-term dynamical SUSY breaking 
 in the context of Dirac gaugino scenario,  
 where the gap equation of D-term has a nontrivial solution \cite{IMaru1, IMaru2}. 
The SUSY breaking vacuum is a necessarily local minimum 
 since we have a trivial solution of the gap equation, i.e. SUSY vacuum, 
 and the vacuum energy in SUSY theories is positive semi-definite. 
It was shown that the lifetime of our SUSY breaking vacuum can be sufficiently long-lived 
 by adjusting the superpotential mass parameter. 
As a phenomenological application, 
we have shown that an observed 126 GeV Higgs mass can be realized by tree level D-term effects 
in a broad range of parameters.

\section*{Acknowledgments}
The work of the author is supported in part by the Grant-in-Aid 
 for Scientific Research from the Ministry of Education, 
 Science and Culture, Japan No. 24540283.


\begin{thebibliography}{10}

\bibitem{DG} 
  P.~Fayet,
  Phys.\ Lett.\ B {\bf 78}, 417 (1978); 
    J.~Polchinski and L.~Susskind,
  Phys.\ Rev.\ D {\bf 26}, 3661 (1982);
  L.~J.~Hall and L.~Randall,
  Nucl.\ Phys.\ B {\bf 352}, 289 (1991);   
  P.~J.~Fox, A.~E.~Nelson and N.~Weiner,
  JHEP {\bf 0208}, 035 (2002). 

\bibitem{IMaru1} 
  H.~Itoyama and N.~Maru,
  Int.\ J.\ Mod.\ Phys.\ A {\bf 27}, 1250159 (2012). 
  
\bibitem{IMaru2}
H.~Itoyama and N.~Maru,
  Phys.\ Rev.\ D {\bf 88}, no. 2, 025012 (2013). 

\bibitem{IMaru3}
H.~Itoyama and N.~Maru,
 Symmetry {\bf 2015} 7(1) 193.  

\end{thebibliography}
\end{document}